\documentclass[10pt,fleqn,twocolumn]{article}
\usepackage[utf8]{inputenc}
\usepackage[english]{babel}
\pdfoutput=1
\usepackage{amsmath,amsfonts,amsthm,mathrsfs}
\usepackage[full]{textcomp}
\usepackage[protrusion=true,expansion=true]{microtype}	
\usepackage{graphicx,color}
\usepackage[a4paper,includeheadfoot,left=1.3cm,right=1.3cm,top=1cm,head=0.5cm,bottom=1.7cm,foot=0.7cm]{geometry}
\usepackage{booktabs}
\usepackage{enumitem}
\usepackage[version=4]{mhchem}
\mhchemoptions{minus-sidebearing-left=0.5em,minus-sidebearing-right=0.5em}
\usepackage{flushend, balance} 

\usepackage[T1]{fontenc}
\usepackage{lmodern}

\usepackage{siunitx}
\DeclareSIUnit\gmol{g\text{-}mol}
\DeclareSIUnit\kgmol{kg\text{-}mol}
\DeclareSIUnit\lbmol{lb\text{-}mol}
\DeclareSIUnit\molar{\mole\per\cubic\deci\metre}
\DeclareSIUnit\Molar{M}
\DeclareSIUnit\torr{torr}
\DeclareSIUnit\micron{\micro\metre}
\DeclareSIUnit\mrad{\milli\rad}
\DeclareSIUnit\gauss{G}
\DeclareSIUnit\rpm{rpm}
\DeclareSIUnit\inch{in}
\DeclareSIUnit\watt{W}
\DeclareSIUnit\ppm{ppm}
\DeclareSIUnit\sccm{sccm}

\usepackage[compress, numbers, super]{natbib}
\setcitestyle{round} 

\usepackage[font=small,labelfont={color=black,bf},textfont={color=black},format=plain,indention=0cm,justification=centerlast,skip=0.1cm]{caption,subfig}
\makeatletter
\renewcommand{\fnum@figure}{\textbf{\small\mbox{Fig.~\thefigure}}}
\renewcommand{\fnum@table}{\textbf{\small\mbox{Table~\thetable}}}
\makeatother

\DeclareRobustCommand{\uppartial}{\text{\rotatebox[origin=t]{19}{\scalebox{0.99}[1]{$\partial$}}}\hspace{-1pt}}

\newcommand{\mytitle}{\textbf{Matrix method for thin film optics}}
\newcommand{\linefindoc}{\color{black}
\vspace{0.5cm}
\centering\rule{0.7\linewidth}{1.2pt}\\%
\vspace{-0.39cm}
\rule{0.5\linewidth}{1.2pt}\\%
\vspace{-0.39cm}
\rule{0.3\linewidth}{1.2pt}%
\vspace{-0.5cm}}
\def\vprod{\boldsymbol{\cdot}} 
\DeclareRobustCommand{\uppartial}{\text{\rotatebox[origin=t]{15}{\scalebox{0.95}[1]{$\partial$}}}\hspace{-1pt}} 

\usepackage[colorlinks=true]{hyperref}
\hypersetup{
pdffitwindow=true,
pdftitle={\mytitle},
pdfauthor={L. N. Acquaroli},
pdfsubject={},
pdfcreator={L.N. Acquaroli},
pdfproducer={L.N. Acquaroli},
pdfkeywords={},
colorlinks={true},
linkcolor={NeonBlue},
citecolor={Cinnabar},
filecolor={Cinnabar},
urlcolor={NeonBlue}
}
\usepackage{url} 

\definecolor{NeonBlue}{rgb}{0.11,0.22,0.73} 
\definecolor{Cinnabar}{rgb}{0.8078,0.0863,0.1255}

\usepackage{titlesec}
\titlelabel{\thetitle. }
\titleformat*{\section}{\centering\large\bfseries}
\titleformat*{\subsection}{\centering\bfseries}

\usepackage{titling}	

\pretitle{\vspace{-40pt} \begin{center} \vspace{5pt} \fontsize{14}{14} \bfseries \color{black} \selectfont }
\title{\mytitle}
\posttitle{\par\end{center}\vspace{-.15cm}}
\preauthor{\vspace{-13pt} \begin{center}
\bigskip \color{black}}
\author{\textbf{Leandro N. Acquaroli}}	
\postauthor{\small \color{black}
\medskip

{Department of Engineering Physics, Ecole Polytechnique Montreal

P.O. Box 6079, Station Centre-Ville, Montreal (QC) H3C 3A7, Canada} 
\vspace{-0.4cm}
\date{\small\today}
\par\end{center}}

\usepackage{fancyhdr}	
\usepackage{lastpage}
\renewcommand{\headrulewidth}{0.0pt}
\renewcommand{\footrulewidth}{0.0pt}

\usepackage{abstract}
\usepackage{indentfirst}

\renewcommand{\thetable}{\Roman{table}}

\begin{document}

\setlength{\belowdisplayskip}{5pt}\setlength{\belowdisplayshortskip}{5pt}
\setlength{\abovedisplayskip}{5pt}\setlength{\abovedisplayshortskip}{5pt}

\columnsep 0.6cm

\renewcommand{\abstractname}{}
\twocolumn[
\maketitle
\vspace{-1.7cm}
\begin{onecolabstract}
Review of a matrix method used in optics of thin films for the calculation of reflectance, transmittance, absorptance, the electric field distribution inside the stack and the photonic dispersion considering the stack as perfect unidimensional crystals ---Distributed Bragg mirrors---. We emphasizes the discussion on transfer matrices and give an alternative approach with scattering matrices for the propagation of light as plane waves through a homogeneous layered system.
\end{onecolabstract}
\vspace{1cm}
]

\fancypagestyle{plain}{%
\fancyhf{} 
\fancyfoot[R]{\footnotesize \thepage\ of \pageref{LastPage}} 
\renewcommand{\headrulewidth}{0pt}
\renewcommand{\footrulewidth}{0pt}}

Thin films are present in diverse applications due to the effective control provided by advanced deposition and electrochemical techniques in the synthesis processes. Functional multilayer stacks offer a broad range of flexibility for their use in optical filters, antireflection coatings and Fabry-Pèrot interferometers~\cite{handbook1,macleod1,bisi1,theiss1}.

The transfer matrix method ---TMM--- reviewed here aims to help predicting the behavior of multilayer thin films structures in a given configuration. The TMM allows analyzing different thin film designs such as single films~\cite{yeh1,monsouri1}, Bragg mirrors ---crystals---, quasycristals ---e.g. Fibonacci or Thue-Morse structures--- according to reflection, transmission, absorption and electromagnetic field distribution~\cite{1742-6596-167-1-012005,ACQUAROLI2010189}. It proved to be useful to calculate the photonic dispersion ---bands structure--- for perfect crystals and to model porosity and thickness gradients~\cite{perez1}. Optofluidic techniques also take advantage of TMM studying the imbibition dynamics inside thin film nanostructures~\cite{doi:10.1021/la104502u,doi:10.1021/la304869y}. We focus on transfer matrices and discuss alternative equations with scattering matrix.

We present the thin film optical theory by steady state Maxwell's equations for the propagation of light through a system of multilayers, assuming the following hypothesis~\cite{knittl1}:
\begin{itemize}[noitemsep,topsep=0mm]
    \item An optically isotropic medium describes the mass of a thin film, characterized by an index of refraction $N\in\mathbb{C}$.
    \item A plane separates two consecutive media with different index of refraction.
    \item The variation of the index of refraction occurs in the direction normal to the multilayer structure ---normal inhomogeneity---.
    \item Two planes define a layer in the propagation axis. The other dimensions of the layer extend to infinity.
    \item The magnitud of the thickness of a layer is in the order of the wavelength of the incident light.
    \item The incident wave is plane, monochromatic and linearly polarized (p or s) respect to the plane of incidence.
\end{itemize}

Consider the following physical aspects that the TMM ignore, but they exists~\cite{knittl1}:
\begin{itemize}[noitemsep,topsep=0mm]
    \item Dispersion of absorption of light caused by polycrystalline structures of evaporated thin films.
    \item The roughness of the substrate and planes ---interfaces--- dividing the layers.
    \item Anisotropy due to internal structures of the material.
    \item Temporal dependence of the index of refraction and thickness ---e.g. aging effects---.
\end{itemize}

\begin{figure}[t]
   \begin{center}
       \includegraphics[scale=0.75]{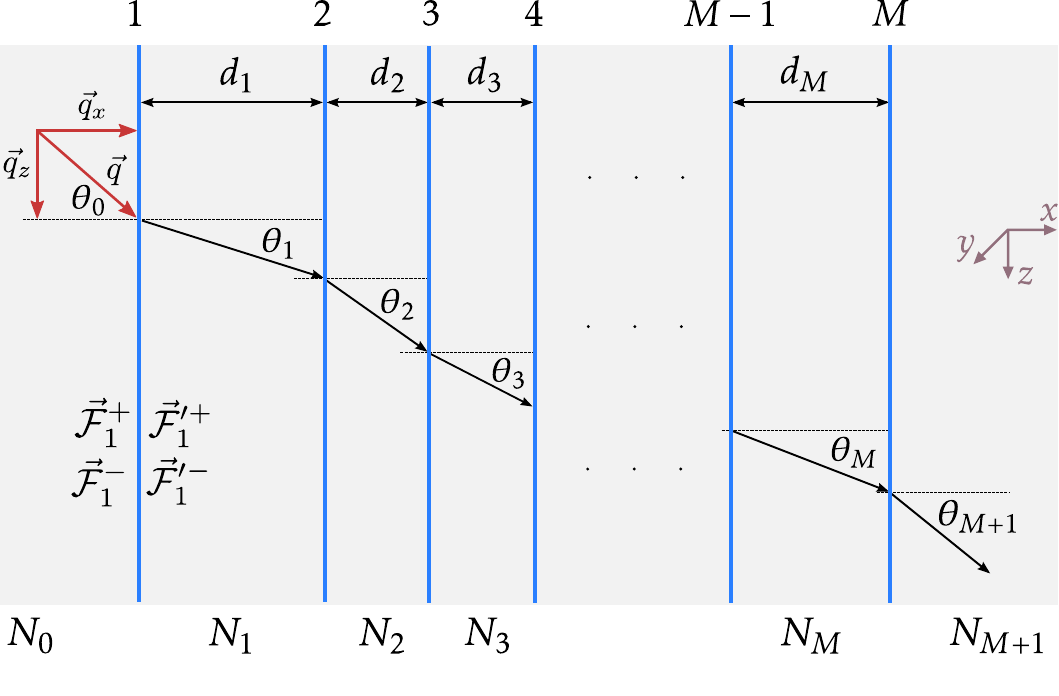}
       \vspace{0.2cm}
       \caption{Scheme of a multilayer stack comprising $j=1,..,M$ layers ---media---, where $d_j$ is the geometrical thickness of each layer, $\theta_j$ is the angle of incidence in the medium normal to the surface, $N_0$ is the index of refraction of the incident medium, $N_j$ is the index of refraction of each layer, and $N_{M+1}$ is the index of refraction of the substrate. $\vec{\mathcal{F}}_{l}^{\pm}$ indicates the vector field ---electric for a p-wave (TM) and magnetic for a s-wave (TE)--- measured before and after crossing interface $l$, that travels towards $\pm x$. The $'$ symbol indicates those quantities located behind ---after crossing the interface--- the optical surfaces~\cite{knittl1}.\hfill{ }}
       \label{f.fig2}
   \end{center}
 \end{figure}

To study the reflection and transmission of the electromagnetic radiation of a multilayer stack, we consider one-unidimensional structures alternating layers with different indexes of refraction in any order ---Fig.~\ref{f.fig2}---. Assuming a wave traveling from $-x$ to $+x$ reflecting at each interface and refracting at each layer of a system composed by $M$ layers, where the wave pass through the last layer experimenting refraction only. These conditions define the dielectric structure as follows:
\begin{equation}\label{eq.1}
    N(x)=\begin{cases}
    N_0,&\quad x<x_1\,, \\
    N_1,&\quad x_1<x<x_2\,,\\
    \cdots&\quad\cdots\\
    N_M,&\quad x_{M-1}<x<x_M\,,\\
    N_{M+1},&\quad x_{M}<x\,,
\end{cases}
\end{equation}

\noindent
where $x_l$ is the position at interface $l$. Maxwell's equations for a linear, non-dispersive, homogeneous, isotropic and without free charges medium read~\cite{saleh}
\begin{align}
    \vec{\nabla}\times\vec{\mathcal{H}}&=\varepsilon\frac{\uppartial\vec{\mathcal{E}}}{\uppartial t},\quad\quad\vec{\nabla}\vprod\vec{\mathcal{H}}=0\,,\nonumber\\
    \vec{\nabla}\times\vec{\mathcal{E}}&=-\mu\frac{\uppartial\vec{\mathcal{H}}}{\uppartial t},\quad\quad\vec{\nabla}\vprod\vec{\mathcal{E}}=0\nonumber\,,
\end{align}

\noindent
where $\varepsilon$ and $\mu$ are the electric permittivity and magnetic permeability of the material, respectively ---for dielectric media, $\mu=1$---. We can write the plane wave solution\footnote{Different authors define it adopting $j = -i$~\cite{knittl1,arnon1}.} to these equations as follow~\cite{cisneros1,jackson1}:
\begin{equation}\label{eq.2}
    \vec{\mathcal{F}}=\vec{\mathscr{F}}\,\exp[i(\vec{q}\vprod\vec{r}-\omega t)]\,,
\end{equation}

 \noindent
where $\vec{\mathscr{F}}$ is the amplitud of the field $\vec{\mathcal{F}}$ ---$\vec{\mathcal{F}}=\vec{\mathcal{E}}$ for p-waves (TM) or $\vec{\mathcal{H}}$ for s-waves (TE)---, $\vec{q}=\hat{{x}}q_{x}+\hat{{z}}q_z$ is the wavevector propagation in the medium and
$\vec{r}=\hat{{x}}x+\hat{{y}}y+\hat{{z}}z$ is the position vector. The wavevector $q_{x,j}^2=q^2-q_z^2=q^2-q^2\sin^2\theta=q^2(1-\sin^2\theta)=q^2\cos^2\theta$, where $q_j=q_0\sqrt{\varepsilon_j\mu_j}=q_0N_j$, and $q_0=\omega/c=2\pi/ \lambda$ is the wavevector in free space~\cite{monsouri1}. For a steady state problem, we can simplify Eq.~\eqref{eq.2} as a linear combination of waves traveling to $-x$ ---regressive waves--- and to $+x$ ---progressive waves---~\cite{cisneros1}:
\begin{align}
    \vec{\mathcal{F}}(x)&=\vec{\mathcal{F}}^+(x)+\vec{\mathcal{F}}^-(x)\\
    &=\vec{\mathscr{F}}_l^+\exp[i q_{x,j}(x-x_l)]+\nonumber\\
    &\quad\,\,\vec{\mathscr{F}}^-_l\exp[-i q_{x,j}(x-x_l)]\label{eq.4}\,.
\end{align}

\begin{figure}[t]
   \begin{center}
       \includegraphics[scale=0.77]{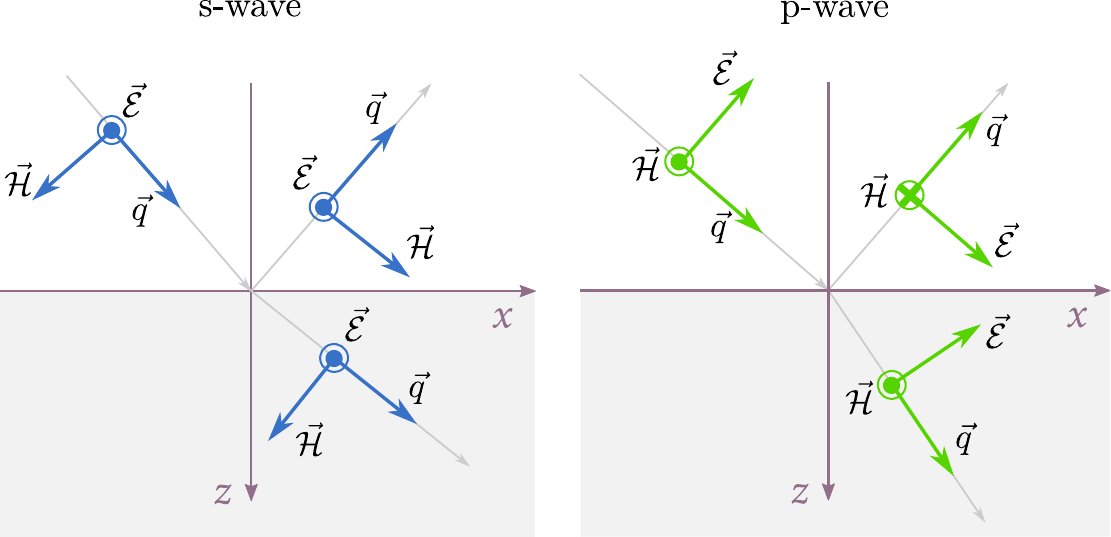}
       \vspace{0.2cm}
       \caption{Orientation of the coordinates system, electromagnetic field and its propagation~\cite{knittl1}.\hfill{ }}
       \label{f.fig3}
   \end{center}
 \end{figure}

We orient the set \{$\vec{\mathcal{E}}$, $\vec{\mathcal{H}}$, $\vec{q}$\} for the incident and reflected waves in such a way that for normal incidence both polarization produce the same results respect to the phase vector $\vec{\mathcal{E}}$~\cite{knittl1}: a change in the axis containing $\vec{\mathcal{H}}$, keeping the axis containing $\vec{\mathcal{E}}$ unchanged after reflection. The orientation of the set remains unaltered in the refracted wave respect to the incident wave ---Fig.~\ref{f.fig3}---.

The optical theory of multilayers consists in repeating the boundary conditions of a simple plane dividing two media, coherently coupling the consecutive boundaries affected by the phase changes applied to the progressive and regressive waves. We can write the boundary conditions taking the tangential components of the electromagnetic fields, $\vec{\mathscr{H}}_{\text{tan}}=\vec{H}$ y $\vec{\mathscr{E}}_{\text{tan}}=\vec{E}$, since they conserve at each side of an interface~\cite{cisneros1,pedrotti1}, employing progressive and regressive waves.

\begin{figure}[t]
   \begin{center}
       \includegraphics[scale=0.78]{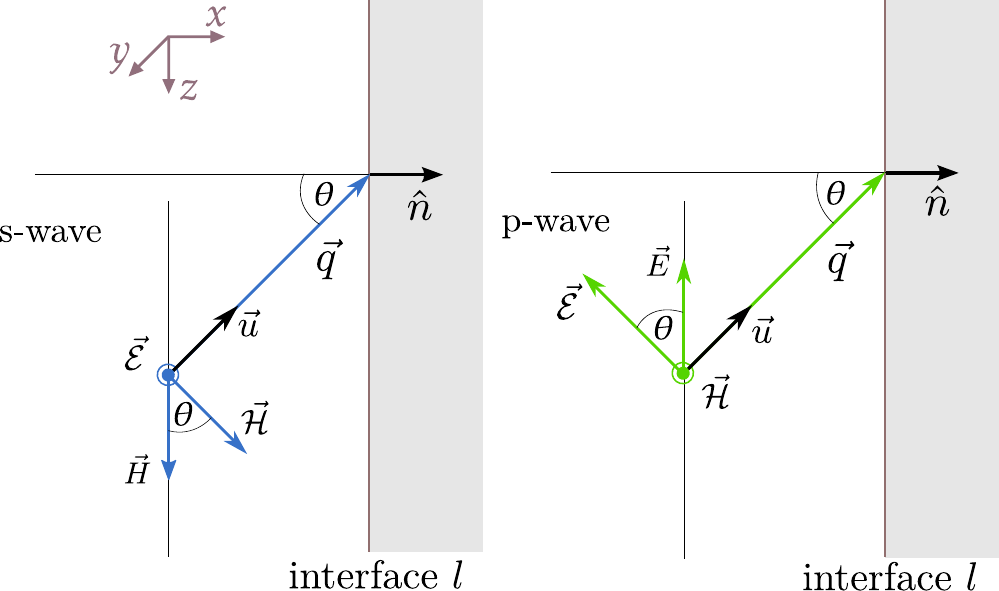}
       \vspace{0.2cm}
       \caption{Scheme of the fields' projections of s- and p-wave. $\hat{{n}}$ is the normal vector. $\hat{n}$ is the vector normal to the surface given by interface $l$, $\vec{q}=\hat{u}q$ is the wavevector, $\hat{u}$ is a unitary vector in the direction of propagation and $\theta$ is the angle of incidence~\cite{knittl1}.\hfill{ }}
       \label{f.fig4}
   \end{center}
 \end{figure}

The boundary conditions for an interface $l$ establish the conservation of the tangential fields at each side of the interface~\cite{knittl1}:
\begin{subequations}\label{eq.7}
    \begin{align}
        \vec{E}_l&=\vec{E}_l^++\vec{E}_l^-=\vec{E}_l^{\prime +}+\vec{E}_l^{\prime -}\label{eq.7a}\\
        \vec{H}_l&=\vec{H}_l^++\vec{H}_l^-=\vec{H}_l^{\prime +}+\vec{H}_l^{\prime -}\,.\label{eq.7b}
\end{align}
\end{subequations}

We define the admittance of a medium $\zeta=\sqrt{\varepsilon/ \mu}$.
According to Fig.~\ref{f.fig4}, the electric field is perpendicular to the surface of incidence, thus, $\vec{\mathcal{E}}=\vec{E}$, while the magnetic field relates to the tangential component as $\hat{{u}}\times\vec{\mathcal{H}}\cos\theta=\hat{{n}}\times\vec{H}$, where $\hat{{u}}$ is the unitary vector in the direction of the wavevector $\vec{q}$. The relation between the tangential components of the fields for a s-wave is then
\begin{equation}\label{eq.add2}
    \hat{{n}}\times\vec{H}=-\zeta\,\vec{E}\,\cos\theta=-s\,\vec{E}\,,
\end{equation}

\noindent
where the parameter $s=\zeta\,\cos\theta$. Taking the cross vector of $\hat{{n}}$ by~\eqref{eq.7b} and using~\eqref{eq.add2} we have:
\begin{align}
    \hat{{n}}\times\vec{H}_l&=\hat{{n}}\times\vec{H}_l^+ +\hat{{n}}\times\vec{H}_l^-\nonumber\\
    &=\hat{{n}}\times\vec{H}_l^{\prime +}+\hat{{n}}\times\vec{H}_l^{\prime -}\nonumber\\ &=s_j\,\vec{E}_l^+-s_j\,\vec{E}_l^-\nonumber\\
    &=s_{j+1}\,\vec{E}_l^{\prime +}-s_{j+1}\,\vec{E}_l^{\prime -}\,.\label{eq.add3}
\end{align}

\noindent
Then, substituting Eq.~\eqref{eq.add3} in \eqref{eq.7b}, the system rearranges as follow:
\begin{subequations}\label{eq.add4}
    \begin{align}
        \vec{E}_l^++\vec{E}_l^-&=\vec{E}_l^{\prime +}+\vec{E}_l^{\prime -}\label{eq.add4a}\\
        s_j\,\vec{E}_l^+-s_j\,\vec{E}_l^-&=s_{j+1}\,\vec{E}_l^{\prime +}-s_{j+1}\,\vec{E}_l^{\prime -}\,.\label{eq.add4b}
    \end{align}
\end{subequations}

\noindent
Performing the same analysis for the p-wave with the magnetic field normal to the surface of incidence, $\vec{\mathcal{H}}=\vec{H}$. Hence, the relation between the electric field with tis tangential component is
$\hat{{n}}\times\vec{E}=\hat{{u}}\times\vec{\mathcal{E}}\cos\theta$. Then,
\begin{equation}\label{eq.add5}
    \hat{{n}}\times\vec{E}=\frac{1}{\zeta}\,\vec{H}\,\cos\theta=s\,\vec{H}\,,
\end{equation}

\noindent
where $s=\cos\theta /\zeta$. Taking the cross vector of $\hat{{n}}$ by~\eqref{eq.7a} and using~\eqref{eq.add5}, the new system reads:
\begin{subequations}\label{eq.add6}
    \begin{align}
        s_j\,\vec{H}_l^+-s_j\,\vec{H}_l^-&=s_{j+1}\,\vec{H}_l^{\prime +}-s_{j+1}\,\vec{H}_l^{\prime -}\label{eq.add6a}\\
        \vec{H}_l^++\vec{H}_l^-&=\vec{H}_l^{\prime +}+\vec{H}_l^{\prime -}\,,\label{eq.add6b}
    \end{align}
\end{subequations}

\noindent
where the negative sign of the last equation is due to that $\vec{H}$ relates to $\vec{E}$ through $-\hat{{n}}$ for the regressive character of the wave.

We define the characteristic matrix of a layer $j$ by
\begin{subequations}\label{eq.8}
    \begin{align}
        \mathit\Gamma_j&=\begin{bmatrix}1&1\\s_j&-s_j\end{bmatrix}\quad \text{s-wave}\,,\label{eq.8a}\\
        \mathit\Gamma_j&=\begin{bmatrix}s_j&-s_j\\1&1\end{bmatrix}\quad \text{p-wave}\,,\label{eq.8b}
    \end{align}
\end{subequations}

\noindent
where $\mathit\Gamma_j\mathit\Gamma_j^{-1}=I$, the identity matrix. Thus, systems~\eqref{eq.add4} and \eqref{eq.add6} in matricial form read:
\begin{equation}\label{eq.9}
    \mathit\Gamma_{j-1}\begin{bmatrix}\vec{F}_l^+\\\vec{F}_l^-\end{bmatrix}=\mathit\Gamma_{j}\begin{bmatrix}\vec{F}_l^{\prime +}\\\vec{F}_l^{\prime -}\end{bmatrix}
\end{equation}

\noindent
or
\begin{equation}\label{eq.10}
    \begin{bmatrix}\vec{F}_l^+\\\vec{F}_j^-\end{bmatrix}=\mathit\Delta_{j-1,j}\begin{bmatrix}\vec{F}_l^{\prime +}\\\vec{F}_l^{\prime -}\end{bmatrix}\,,\quad \mathit\Delta_{j-1,j}=\mathit\Gamma_{j-1}^{-1}\mathit\Gamma_{j}\,.
\end{equation}

\noindent
Equation~\eqref{eq.10} describes the relation between the incoming and outgoing fields at the interface $j$, where $\mathit\Delta$ is the transfer matrix~\cite{cisneros1} ---also called transformation or refraction matrix~\cite{knittl1}---, that satisfies the relation $\det\{\mathit\Delta_{j-1,j}\}=s_{j}/s_{j-1}$. After crossing the interface $l$ the wave propagates certain distance until the next interface $l+1$. The distance between these two consecutive interfaces equals the thickness of the layer $j$, $d_j$. The progressive and regressive waves, according to \eqref{eq.4}, are:
\begin{subequations}\label{eq.11}
    \begin{align}
        \vec{F}_l^+(x_l=0)&=\vec{\mathscr{F}}_l^+\label{eq.11a}\\
        \vec{F}_l^-(x_l=0)&=\vec{\mathscr{F}}_l^-\label{eq.11b}\\
        \vec{F}_{l+1}^+(x_{l+1}=d_j)&=\vec{\mathscr{F}}_{l+1}^+\exp(i q_{x,j} d_j)\label{eq.11c}\\
        \vec{F}_{l+1}^-(x_{l+1}=d_j)&=\vec{\mathscr{F}}_{l+1}^-\exp(-i q_{x,j} d_j)\label{eq.11d}\,.
\end{align}
\end{subequations}

\noindent
Combining \eqref{eq.11a} with \eqref{eq.11c} and \eqref{eq.11b} with \eqref{eq.11d} we have:
\begin{align}
    \vec{F}_l^+(0)&=\vec{F}_{l+1}^+(d)\exp(-i q_{x,j} d_j)\\
    \vec{F}_l^-(0)&=\vec{F}_{l+1}^-(d)\exp(i q_{x,j} d_j)\,.
\end{align}

\noindent
A general expression results writing the previous equations in matricial form:
\begin{equation}\label{eq.12c}
    \begin{bmatrix}\vec{F}^{\prime +}_l\\\vec{F}^{\prime -}_l\end{bmatrix}= \begin{bmatrix}e^{-i\varphi_j}&0\\0&e^{i\varphi_j}\end{bmatrix} \begin{bmatrix}\vec{F}^+_{l+1}\\\vec{F}^-_{l+1}\end{bmatrix}= \mathit\Upsilon_j\begin{bmatrix}\vec{F}^+_{l+1}\\\vec{F}^-_{l+1}\end{bmatrix}\,,
\end{equation}

\noindent
where
\begin{equation}\label{eq.12}
    \varphi_j=q_x d_j=\frac{2\pi}{\lambda}N_j d_j\cos\theta_j
\end{equation}

\noindent
is the phase shift angle experimented by the wave after crossing the layer $j$. $\mathit\Upsilon_j$ is the propagation~\cite{monsouri1} or phase~\cite{knittl1} matrix, which is unimodular: $\det\{\mathit\Upsilon_j\}=1$.

Merging the matrices relating the fields at both sides of the interface and the propagation through a layer, we can compute the total matrix of a multilayer structure, using Eqs.~\eqref{eq.10} and \eqref{eq.11} for a total number of $M$ layers~\cite{knittl1}:
\begin{multline}\label{eq.13}
    \begin{bmatrix}\vec{F}_1^+\\\vec{F}_1^-\end{bmatrix}=\mathit\Delta_{0,1}\mathit\Upsilon_{1}\mathit\Delta_{1,2}\mathit\Upsilon_{2}\cdots\\
\cdots\mathit\Delta_{M-1,M}\mathit\Upsilon_{M}\mathit\Delta_{M,M+1}\begin{bmatrix}\vec{F}_{M+1}^{\prime +}\\\vec{F}_{M+1}^{\prime -}\end{bmatrix}\,.
\end{multline}

\noindent
Taking the product of the r.h.s. of the last expression previous to the column vector, we define the total transfer matrix of the system, $\mathit\Omega$, as follow:
\begin{equation}
    \begin{bmatrix}\vec{F}_1^+\\\vec{F}_1^-\end{bmatrix}=\prod_{j=1}^{j=M}\mathit\Delta_{j-1,j}\mathit\Upsilon_{j}\mathit\Delta_{j,j+1} \begin{bmatrix}\vec{F}_{M+1}^{\prime +}\\ \vec{F}_{M+1}^{\prime -}\end{bmatrix}=\mathit\Omega\begin{bmatrix}\vec{F}_{M+1}^{\prime +}\\ \vec{F}_{M+1}^{\prime -}\end{bmatrix}\,.\label{eq.14}
\end{equation}

\noindent
The matrix $\mathit\Omega$ relates the tangential components of the fields $+$ and $-$ at the extremes of the multilayer. We define the interference matrix $\mathit\Phi_j$ for both polarization as~\cite{monsouri1,cisneros1,luca1}
\begin{equation}\label{eq.15}
    \mathit\Phi_j=\mathit\Gamma_j\mathit\Upsilon_j\mathit\Gamma_j^{-1}=\begin{bmatrix}\cos\varphi_j&-(i/s_j)\sin\varphi_j\\-i s_j \sin\varphi_j&\cos\varphi_j\end{bmatrix}\,.
\end{equation}

\noindent
$\mathit\Phi$ is unimodular and it relates to the transfer matrix of the system $\mathit\Omega$ as follow~\cite{knittl1,monsouri1}:
\begin{equation}\label{eq.16}
    \mathit\Omega=\mathit\Gamma_0^{-1}\left[\prod_{j=1}^{j=M}\mathit\Phi_j\right]\mathit\Gamma_{M}=\mathit\Gamma_0^{-1}\mathit\Psi\mathit\Gamma_{M+1}\,.
\end{equation}

\noindent
where $\mathit\Psi$ is the interference matrix of the system and establish the transformation of the incoming and outgonig tangential total fields in the system,
\begin{equation}\label{eq.17}
    \begin{bmatrix}\vec{E}_1\\\vec{H}_1\end{bmatrix}=\mathit\Psi\begin{bmatrix}\vec{E}_{M+1}\\\vec{H}_{M+1}\end{bmatrix}\,.
\end{equation}

We can further use the matrix theory described until now to calculate the reflection, transmission and absorption spectra of the multilayer structure in terms of the transfer and the interference matrices. Expanding Eq.~\eqref{eq.14}
\begin{subequations}\label{eq.18}
    \begin{align}
        \vec{F}_1^+&=\omega_{1,1}\vec{F}_{M+1}^{\prime +}+\omega_{1,2}\vec{F}_{M+1}^{\prime -}\label{eq.18a}\\
        \vec{F}_1^-&=\omega_{2,1}\vec{F}_{M+1}^{\prime +}+\omega_{2,2}\vec{F}_{M+1}^{\prime -}\label{eq.18b}\,,
\end{align}
\end{subequations}

\noindent
the reflection $\tilde{r}$ and transmission $\tilde{t}$ Fresnell coefficients for both directions of incident light can be determined. Consider first the progressive waves, $\vec{F}_{M+1}^{\prime -}=0$, i.e. after crossing the last layer, the wave does not undergoes any reflection. Then,
\begin{align}
    \tilde{r}^+&=\frac{\vec{F}_1^-}{\vec{F}_1^+}=\frac{\omega_{2,1}}{\omega_{1,1}}\label{eq.19}\\
    \tilde{t}^+&=\frac{\vec{F}_{M+1}^+}{\vec{F}_1^+}=\frac{1}{\omega_{1,1}}\label{eq.20}\,.
\end{align}

\noindent
For regressive waves, $\vec{F}_1^+=0$, then
\begin{align}
    \tilde{r}^-&=\frac{\vec{F}_{M+1}^{\prime -}}{\vec{F}_{M+1}^{\prime +}}=-\frac{\omega_{1,2}}{\omega_{1,1}}\label{eq.22}\\
    \tilde{t}^-&=\frac{\vec{F}_1^-}{\vec{F}_{M+1}^{\prime -}}=\frac{1}{\omega_{1,1}}\det\{\mathit\Omega\}\label{eq.23}\,.
\end{align}

$\mathit\Omega$ results from the product of $\mathit\Delta$ and $\mathit\Upsilon$, thus, $\det\{\Omega\}=s_{M+1}s_0^{-1}$, leading to the important relation~\cite{knittl1}:
\begin{equation}\label{eq.24}
    \tilde{t}^-=\frac{s_{M+1}}{s_0}\tilde{t}^+\,.
\end{equation}

According to Eq.~\eqref{eq.16} we can relate the elements of $\mathit\Omega$ with those of $\mathit\Psi$
\begin{subequations}\label{eq.25}
    \begin{align}\omega_{1,1}&=\frac{1}{2}(s_0\psi_{1,1}+\psi_{2,1}+\nonumber\\
        &\quad\quad\quad\quad s_0s_{M+1}\psi_{1,2}+s_{M+1}\psi_{2,2})\label{eq.25a}\\
        \omega_{1,2}&=\frac{1}{2}(s_0\psi_{1,1}+\psi_{2,1}-\nonumber\\
        &\quad\quad\quad\quad s_0s_{M+1}\omega_{1,2}-s_{M+1}\psi_{2,2})\label{eq.25b}\\
        \omega_{2,1}&=\frac{1}{2}(s_0\psi_{1,1}-\psi_{2,1}+\nonumber\\
        &\quad\quad\quad\quad s_0s_{M+1}\psi_{1,2}-s_{M+1}\psi_{2,2})\label{eq.25c}\\
        \omega_{2,2}&=\frac{1}{2}(s_0\psi_{1,1}-\psi_{2,1}-\nonumber\\
        &\quad\quad\quad\quad s_0s_{M+1}\psi_{1,2}-s_{M+1}\psi_{2,2})\,.\label{eq.25d}
\end{align}
\end{subequations}

\noindent
and then calculate the reflection and transmission coefficients as follow~\cite{knittl1,pedrotti1}:
\begin{align}
    \tilde{r}^+&=\frac{s_0\psi_{1,1}-\psi_{2,1}+s_0s_{M+1}\psi_{1,2}-s_{M+1}\psi_{2,2}}{s_0\psi_{1,1}+s_0s_{M+1}\psi_{1,2}+\psi_{2,1}+s_{M+1}\psi_{2,2}}\,,\label{eq.26}\quad\\
    \tilde{t}^+&=\frac{2}{s_0\psi_{1,1}+s_0s_{M+1}\psi_{1,2}+\psi_{2,1}+s_{M+1}\psi_{2,2}}\label{eq.27}\,.
\end{align}

\noindent
The expression for the reflectance and transmittance from the coefficients derived are:
\begin{align}
    R&=\tilde{r}^\pm(\tilde{r}^\pm)^*\label{eq.28}\\
    T&=s_0s_{M+1}\tilde{t}^\pm(\tilde{t}^\pm)^*\,,\label{eq.29}
\end{align}

\noindent
where $^*$ denotes the complex conjugate. Cisneros \textit{et. al} explain that \eqref{eq.29} is valid when the last medium is non-absorbent~\cite{cisneros1}, although, a more general expression is proposed taking the real part, $\Re{[s_{M+1}]}$~\cite{macleod1}. We do not include the absorptance in terms of the matrix elements, as it is simply calculated by $A=1-T-R$~\cite{cisneros1}.

There exists a direct relation between the absorption and the intensity of the field at any point inside the multilayer structures. Computing the electromagnetic field distribution allows to analyze important effects such as the damage induced by a laser radiation on the layers, in which the absorption transforms into incident heat energy~\cite{arnon1,apfel1,apfel2,demichelis1,doi:10.1117/12.2030380}. The enhancement of the field inside Fabry-Pèrot type cavities provoque an increase in the FTIR and Raman signals, which is useful to study intrinsically weak vibrational modes~\cite{mattei1}.

We define normalized field distribution as follow~\cite{arnon1}
\begin{equation}\label{eq.30a}
    I=\frac{|\vec{F}(x)|^2}{|\vec{F}_1^+|^2}\,,
\end{equation}

\noindent
where $\vec{F}(x)$ is the total field at the position $x$ inside the multilayer, and $\vec{F}_1^+$ is the incident field of the progressive wave, where $x=0$ is the origin of the first layer in the stack. Since the wave travels towards $+x$, Eq.~\eqref{eq.18a} establish that $\vec{F}_{M+1}^{\prime -}=0$, then $\vec{F}_1^+=\omega_{1,1}\vec{F}_{M+1}^{\prime +}$ for the first interface. The next step is to calculate the field as a function of the position $x$. A simple approach to do this is taking the product between the total matrix $\mathit\Psi$ by $\mathit\Xi=\mathit\Phi^{-1}$:
\begin{equation*}
    \mathit\Xi_\ell=\begin{bmatrix}\cos\vartheta_\ell&(i/s_\ell)\,\sin\vartheta_\ell\\i\,s_\ell\,\sin\vartheta_\ell&\cos\vartheta_\ell\end{bmatrix}\,,
\end{equation*}

\noindent
where the elements varies for each position inside the multilayer through the phase shift angle $\vartheta_\ell$:
\begin{equation*}
    \vartheta_\ell=\frac{1}{h}\,\left(\frac{2\,\pi}{\lambda}\,N_\ell\,d_\ell\,\cos\theta_\ell\right)\,.
\end{equation*}

\noindent
The constant $h$ is the number of times we divide the phase shift angle to compute the electromagnetic field at the position $x\in[0,h\cdot M]$. Taking the product of $\mathit\Xi$ times the total $\mathit\Psi$,
\begin{equation*}
    G(x)=\left(\prod_\ell^x\,\Xi_\ell\right)\,\mathit\Psi\,,
\end{equation*}

\noindent
determines the field at each position through
\begin{equation*}
    \vec{F}(x)=[g_{1,1}(x)+g_{1,2}(x)\,s_{M+1}]\vec{F}_{M+1}^{\prime +}\,,
\end{equation*}

\noindent
where $g$ are the elements of the matrix $G$. The intensity ratio ---Eq.~\eqref{eq.30a}--- takes the final form:
\begin{equation}\label{eq.36}
    I(x)=\frac{|g_{1,1}(x)+g_{1,2}(x)\,s_{M+1}|^2}{|\gamma_{1,1}|^2}\,.
\end{equation}

A wave in a periodic system travels similarly to electrons in a crystalline solid. Hence, we can borrow the mathematical formulation for the band theory in solids and apply it to the electromagnetic propagation in periodic media, along with the concepts of Bloch waves, Brillouin zone and band-gaps. A binary ---alternates two media with different index of refraction--- periodic system resembles an unidimensional lattice invariant to translation operation. The relation between the waves amplitudes in a unit cell of a periodic multilayer is~\cite{perez1}:
\begin{align}
    \begin{bmatrix}\vec{F}_l^+\\\vec{F}_l^-\end{bmatrix}&=\Delta_{j,j+1}\Upsilon_{j+1}\Delta_{j+1,j+2}\Upsilon_{j+2}\begin{bmatrix}\vec{F}_{l+2}^+\\\vec{F}_{l+2}^-\end{bmatrix}\nonumber\\
        &=U\begin{bmatrix}\vec{F}_{l+2}^+\\\vec{F}_{l+2}^-\end{bmatrix}\,,\label{eq.37}
\end{align}

\noindent
where $U$ is the translation operator in the unit cell. According to Bloch's ---Floquet--- theorem a wave propagates in a periodic system in the form of~\cite{arnon1}
\begin{equation}\label{eq.38}
    \vec{F}_K(x,z)=\vec{F}_K(x)\exp{(i Kx)}\,\exp{(i q_z z)}
\end{equation}

\noindent
where $\vec{F}_K$ is periodic with period $\Lambda$, where $\Lambda$ ---unit cell--- results from adding the thicknesses of the two layers with different indexes of refraction gives the period:
\begin{equation}\label{eq.39}
    \vec{F}_K(x+\Lambda)=\vec{F}_K(x)\,.
\end{equation}

\noindent
The quantity to determine is the constant $K$, the Bloch wavevector. Rewriting condition~\eqref{eq.39} in terms of Eq.~\eqref{eq.4}, results in
\begin{equation}\label{eq.40}
    \begin{bmatrix}\vec{F}_l^+\\\vec{F}_l^-\end{bmatrix}=\exp{(-i K\Lambda)}\begin{bmatrix}\vec{F}_{l+2}^+\\\vec{F}_{l+2}^-\end{bmatrix}\,.
\end{equation}

\noindent
Combining Eqs.~\eqref{eq.39} and~\eqref{eq.40} we note that the Bloch wave satisfies the following eigenvalue equation:
\begin{equation}\label{eq.41}
    U\begin{bmatrix}\vec{F}_l^+\\\vec{F}_l^-\end{bmatrix}=\exp{(i K\Lambda)}\begin{bmatrix}\vec{F}_l^+\\\vec{F}_l^-\end{bmatrix}\,.
\end{equation}

\noindent
The phase factor is the eigenvalue of the translation operator $U$, given by
\begin{multline}\label{eq.42}
    e^{\pm i K\Lambda}=\frac{1}{2}(u_{1,1}+u_{2,2})\pm\\
    i\left\{1-\left[\frac{1}{2}(u_{1,1}+u_{2,2})\right]^2\right\}^{1/2}\,.
\end{multline}

\noindent
Equation~\eqref{eq.41} allows to calculate the corresponding eigenvectors as follow,
\begin{equation}\label{eq.43}
    \begin{bmatrix}\vec{F}_1^+\\\vec{F}_1^-\end{bmatrix}=\begin{bmatrix}u_{1,2}\\\exp{(-i K\Lambda)}-u_{1,1}\end{bmatrix}\,.
\end{equation}

\noindent
multiplied by an arbitrary constant~\cite{arnon1}. The Bloch waves~\eqref{eq.43} are the eigenvectors of the translation operator with eigenvalues $\exp{(\pm i K\Lambda)}$ given by~\eqref{eq.42}. Both eigenvalues are inverse to each other since the matrix $U$ is unimodular. Equation~\eqref{eq.42} describes the relation dispersion between the frequency $\varpi$, the wavevector $q_z$ and the Bloch vector $K$ for the Bloch wave function:
\begin{align}\label{eq.44}
    K(\varpi,q_z)&=\frac{1}{\Lambda}\arccos\left[\text{tr}(U)\right]\nonumber\\
    &=\frac{1}{\Lambda}\arccos\left[\frac{1}{2}(u_{1,1}+u_{2,2})\right]\,.
\end{align}

\noindent
Three regimes arise from Eq.~\eqref{eq.44}. When $|1/2(u_{1,1}+u_{2,2})|<1$, $K$ is real and the Bloch wave propagates, while if $|1/2(u_{1,1}+u_{2,2})| >1$, then $K=m\pi / \Lambda+ i K_i$ has an imaginary component and the Bloch wave is evanescent. These last waves represent forbidden band-gaps in a periodic system. The edges of the bands locate in the regime where $|1/2(u_{1,1}+u_{2,2})|=1$. An alternative expression for the dispersion relation expanding~\eqref{eq.44} results as follow,
\begin{multline}
    \cos(K\Lambda)=\frac{1}{2}\left[2\cos\varphi_1\cos\varphi_2\right.\\
    \left. -\frac{(s_1^2+s_2^2)}{s_1s_2}\sin\varphi_1\sin\varphi_2\right]\,,\label{eq.45}
\end{multline}

\noindent
where $\varphi$ is the phase shift angle from~\eqref{eq.12}.

\begin{figure}[t]
   \begin{center}
       \includegraphics[scale=0.9]{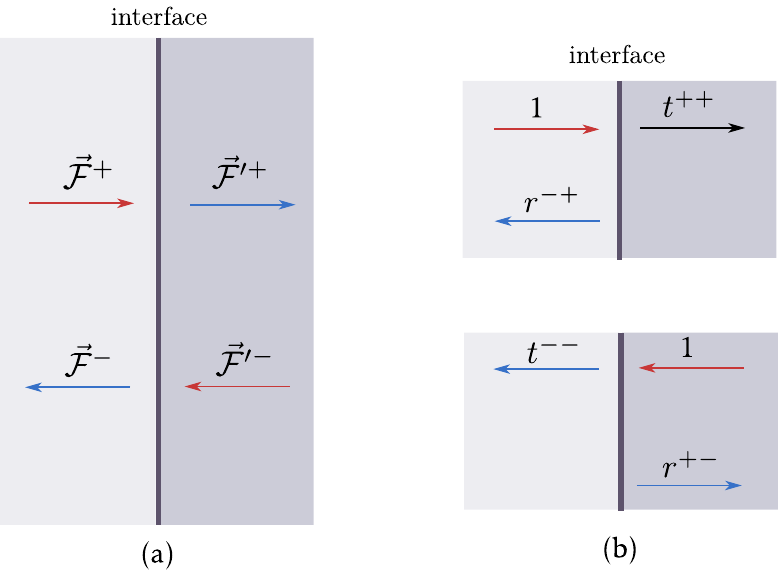}
       \vspace{0.2cm}
       \caption{(a)~Incoming and outgoing waves at an interface. (b)~Scattering of incoming waves in terms of the scattering coefficients $t^{++}$, $t^{--}$, $r^{-+}$ and $r^{+-}$.\hfill{ }}
       \label{f.mmt1}
   \end{center}
 \end{figure}

An alternative approach to the TMM formalism which is the scattering matrices method, defined as $X$ matrices~\cite{saleh,pendry}. The base of this method is to express the outgoing waves from a scattering center as a function of the incoming waves ---Fig.~\ref{f.mmt1}---. The scattering relations require the amplitudes to satisfy
\begin{align}\label{eq.mmt1}
    \vec{F}_l^{\prime +}&=t^{++}\vec{F}_l^{+}+r^{+-}\vec{F}_l^{\prime -}\\
    \vec{F}_l^{-}&=r^{-+}\vec{F}_l^{+}+t^{--}\vec{F}_l^{\prime -}\,.
\end{align}

\noindent
In matricial form the last equation reads:
\begin{equation}\label{eq.mmt2}
    \begin{bmatrix}1&-r^{+-}\\0&t^{--}\end{bmatrix}\begin{bmatrix}\vec{F}_l^{\prime +}\\\vec{F}_l^{\prime -}\end{bmatrix}=\begin{bmatrix}t^{++}&0\\-r^{-+}&1\end{bmatrix}\begin{bmatrix}\vec{F}_l^{+}\\\vec{F}_l^{-}\end{bmatrix}\,.
\end{equation}

\noindent
Inverting the matrix on the left of Eq.~\eqref{eq.mmt2}, results
\begin{align}\label{eq.mmt3}
    \begin{bmatrix}\vec{F}_l^{\prime +}\\\vec{F}_l^{\prime -}\end{bmatrix}&=\begin{bmatrix}t^{++}-r^{+-}(t^{--})^{-1}r^{-+}&r^{+-}(t^{--})^{-1}\\-(t^{--})^{-1}r^{-+}&(t^{--})^{-1}\end{bmatrix}\begin{bmatrix}\vec{F}_l^{+}\\\vec{F}_l^{-}\end{bmatrix}\\
    &=X\begin{bmatrix}\vec{F}_l^{+}\\\vec{F}_l^{-}\end{bmatrix}\,.
\end{align}

\noindent
The expressions relating the transfer matrix with the scattering matrix at an interface results from the combination of Eqs.~\eqref{eq.10} and \eqref{eq.mmt3}:
\begin{align}
    \mathit\Delta&=\begin{bmatrix}t^{++}-r^{+-}(t^{--})^{-1}r^{-+}&r^{+-}(t^{--})^{-1}\\-(t^{--})^{-1}r^{-+}&(t^{--})^{-1}\end{bmatrix}^{-1}\\
    X&=\begin{bmatrix}\delta_{1,1}-\delta_{1,2}\delta_{2,1}\delta_{2,2}^{-1}&\delta_{1,2}\delta_{2,2}^{-1}\\-\delta_{2,1}\delta_{2,2}^{-1}&\delta_{2,2}^{-1}\end{bmatrix}^{-1}\,.
\end{align}

\noindent
For a wave crossing an homogeneous layer the scattering matrix turns out to be:
\begin{equation}
    X=\begin{bmatrix}e^{-i\varphi}&0\\0&e^{-i\varphi}\end{bmatrix}\,,
\end{equation}

\noindent
where $\varphi$ is the phase shift angle. Notice that this equation differs from that expressed by $\mathit\Upsilon$ in Eq.~\eqref{eq.12c}.

We can summarize the main characteristics of the TMM as follow:
\begin{itemize}[noitemsep,topsep=0mm]
    \item Efficiently calculates the optical spectra of arbitrary ordered multilayer systems.
    \item Handle complex index of refraction denoting the gaining or absorption for cases of negative or positive index of refraction. When the index is real it ideally behaves without dissipation of energy ---lossless material---.
    \item The thicknesses of the layers can take any value. Although, we can expect incoherence effects.
    \item Suitable to calculate the distribution of the electric field throughout a multilayer stack.
    \item Assumes the plane perpendicular to the direction of propagation to be infinite, implicating that each layer extends infinitely in other dimensions. The incident and outgoing ---substrate--- media are semi-infinite.
    \item Calculates the fields in the structure propagating from one layer to the next one by matrix relations, making the computational cost dependable on the number of layers.
    \item Limited to waves traveling continuously without pulses of propagation, where finite difference techniques becomes useful.
    \item Handle dispersion relations for perfect crystals or periodic binary systems.
\end{itemize}

\linefindoc
\bibliographystyle{unsrt}

\begin{thebibliography}{10}

\bibitem{handbook1}
J.~A. Dobrowolski.
\newblock Fundamentals, techniques, and design.
\newblock In {\em Handbook of Optics}, volume~1, chapter~42. McGraw-Hill, New
  York, 2 edition, 1994.

\bibitem{macleod1}
H.~A. Macleod.
\newblock {\em Thin -Film Optical Filters}.
\newblock Institute of Physics Publishing, 3 edition, 2001.

\bibitem{bisi1}
O.~Bisi, E.~Ossicini, and L.~Pavesi.
\newblock Porous silicon: A quantum sponge structure for silicon based
  optoelectronics.
\newblock {\em Surface Science Reports}, 38:1--126, 2000.

\bibitem{theiss1}
W.~Thei\ss.
\newblock Optical properties of porous silicon.
\newblock {\em Surface Science Reports}, 29(3--4):91--192, 1997.

\bibitem{yeh1}
P.~Yeh, A.~Yariv, and C.~S. Hong.
\newblock Electromagnetic propagation in periodic stratified media. i. general
  theory.
\newblock {\em Journal of the Optical Society of America}, 67(4):423, 1997.

\bibitem{monsouri1}
J.~A. Monsouri, R.~A. Depine, and E.~Silvestre.
\newblock Porous silicon: A quantum sponge structure for silicon based
  optoelectronics.
\newblock {\em Journal of the European Optical Society - Rapid Publications},
  2:07002, 2007.

\bibitem{1742-6596-167-1-012005}
R.~Urteaga, O.~Marín, L.~N. Acquaroli, D.~Comedi, J.~A. Schmidt, and R.~R.
  Koropecki.
\newblock Enhanced photoconductivity and fine response tuning in nanostructured
  porous silicon microcavities.
\newblock {\em Journal of Physics: Conference Series}, 167(1):012005, 2009.

\bibitem{ACQUAROLI2010189}
L.~N. Acquaroli, R.~Urteaga, and R.~R. Koropecki.
\newblock Innovative design for optical porous silicon gas sensor.
\newblock {\em Sensors and Actuators B: Chemical}, 149(1):189 -- 193, 2010.

\bibitem{perez1}
E.~X. Pérez.
\newblock {\em Design, fabrication and characterization of porous silicon
  multilayer optical devices}.
\newblock PhD thesis, Universitat Rovira I Virgili, Tarragona, 2007.

\bibitem{doi:10.1021/la104502u}
L.~N. Acquaroli, R.~Urteaga, C.~L.~A. Berli, and R.~R. Koropecki.
\newblock Capillary filling in nanostructured porous silicon.
\newblock {\em Langmuir}, 27(5):2067--2072, 2011.

\bibitem{doi:10.1021/la304869y}
R.~Urteaga, L.~N. Acquaroli, R.~R. Koropecki, A.~Santos, M.~Alba,
  J.~Pallar\`{e}s, L.~F. Marsal, and C.~L.~A. Berli.
\newblock Optofluidic characterization of nanoporous membranes.
\newblock {\em Langmuir}, 29(8):2784--2789, 2013.

\bibitem{knittl1}
Z.~Knittl.
\newblock {\em Optics of Thin Films (An Optical Multilayer Theory)}.
\newblock John Wiley \& Sons, Czechoslovakia, 1976.

\bibitem{saleh}
B.~E.~A. Saleh and M.~C. Teich.
\newblock {\em Fundamentals of photonics}.
\newblock John Wiley \& Sons, 2 edition, 2007.

\bibitem{arnon1}
O.~Arnon and P.~Baumeister.
\newblock Electric field distribution and the reduction of laser damage in
  multilayers.
\newblock {\em Applied Optics}, 19(11):1853, 1980.

\bibitem{cisneros1}
J.~I. Cisneros.
\newblock {\em Ondas Eletromagnéticas. Fundamentos e aplicaç\~{o}es}.
\newblock Editora da UNICAMP, Campinas, SP Brasil, 2001.

\bibitem{jackson1}
J.~D. Jackson.
\newblock {\em Classical Electrodynamics}.
\newblock John Wiley \& Sons, 3 edition, 1998.

\bibitem{pedrotti1}
F.~J. Pedrotti and L.~S. Pedrotti.
\newblock {\em Introduction to Optics}.
\newblock Prentice Hall, USA, 2 edition, 1992.

\bibitem{luca1}
L.~Plattner.
\newblock {\em A Study in Biomimetics: Nanometer-scale, high-efficiency,
  dielectric diffractive structures on the wings of butterflies and in the
  silicon chip factory}.
\newblock PhD thesis, University of Southampton, 2003.

\bibitem{apfel1}
J.~H. Apfel.
\newblock Electric fields in multilayers at oblique incidence.
\newblock {\em Applied Optics}, 15(10):2339, 1976.

\bibitem{apfel2}
J.~H. Apfel.
\newblock Optical coating design with reduced electric field intensity.
\newblock {\em Applied Optics}, 16(7):1880, 1977.

\bibitem{demichelis1}
F.~Demichelis, E.~Mezzetti-Minetti, and E.~Tresso.
\newblock Optimization of optical parameters and electric field distribution in
  multilayers.
\newblock {\em Applied Optics}, 23(1):165, 1984.

\bibitem{doi:10.1117/12.2030380}
D.~Patel, D.~Schiltz, P.~F. Langton, L.~Emmert, L.~N. Acquaroli, C.~Baumgarten,
  B.~Reagan, J.~J. Rocca, W.~Rudolph, A.~Markosyan, R.~R. Route, M.~Fejer, and
  C.~S. Menoni.
\newblock Improvements in the laser damage behavior of {Ta$_2$O$_5$/SiO$_2$}
  interference coatings by modification of the top layer design.
\newblock {\em Proc. SPIE}, 8885:8885--1 -- 8885--5, 2013.

\bibitem{mattei1}
G.~Mattei, G.~Marucci, and V.~A. Yakovlev.
\newblock Splitting of porous silicon microcavity mode due to the interaction
  with si–h vibrations.
\newblock {\em Materials Science and Engineering B}, 51(1--3):158, 1998.

\bibitem{pendry}
J.~B. Pendry.
\newblock Waves in 1d disordered systems.
\newblock {\em Advances in physics}, 43(4):461--542, 1995.

\end{thebibliography}

\end{document}